\documentclass[twocolumn]{article}
\usepackage[utf8]{inputenc}
\usepackage{geometry}
\usepackage{graphicx} 
\usepackage{pgfplots} 
\pgfplotsset{compat=1.16}
\usepackage{filecontents} 
\usepackage{tikz} 
\usepackage{amsmath} 
\usepackage{float}
\usepackage{amssymb}
\usepackage[scale=2]{ccicons}

\usepackage{hyperref}

\title{A new adaptive pricing framework for perpetual protocols using liquidity curves and on-chain oracles}

\author{
    Chester Bella\\
    cb@parifi.org
\and
    Danny Boahen\\
    danny@parifi.org
\and
    Sudeep Biswas\\
    sudeep@parifi.org
}
\date{\vspace{0.5cm}April 2023}

\makeatletter
\renewcommand{\@maketitle}{%
\newpage
 \null
 \vskip 2em%
 \begin{center}%
  \let \footnote \thanks
  {\LARGE \@title \par}%
  \vskip 3em
  {\large
   \lineskip .5em%
   \begin{tabular}[t]{c}%
    \@author
   \end{tabular}\par}%
  \vskip 1em%
  {\large \@date}%
 \end{center}%
 \par
 \vskip 1.5em}
\makeatother

\begin{document}

\maketitle

\begin{abstract}
This whitepaper introduces an innovative mechanism for pricing perpetual contracts and quoting fees to traders based on current market conditions. The approach employs liquidity curves and on-chain oracles to establish a new adaptive pricing framework that considers various factors, ensuring pricing stability and predictability. The framework utilizes parabolic and sigmoid functions to quote prices and fees, accounting for liquidity, active long and short positions, and utilization. This whitepaper provides a detailed explanation of how the adaptive pricing framework, in conjunction with liquidity curves, operates through mathematical modeling and compares it to existing solutions. Furthermore, we explore additional features that enhance the overall efficiency of the decentralized perpetual protocol.\end{abstract}

\section{Introduction}

Perpetual futures contracts, a relatively recent innovation in derivatives trading, are gaining traction in the cryptocurrency landscape. Sharing similarities with conventional futures contracts, they allow traders to trade on an asset's future value without actually possessing the asset. A key distinction lies in the fact that perpetual futures contracts don't have an expiration date. This feature enables traders to maintain their positions indefinitely, as long as they uphold a healthy margin ratio. In turn, this provides traders with enhanced flexibility and the capacity to effectively manage their risk exposure.
Several approaches to implement perpetual contracts have emerged over time to solve inefficiencies, each with its own trade-offs. There are two main approaches to market making in perpetual contracts: Order book-based and Automated Market Maker (AMM) based systems. In order book-based market making, traders place buy and sell orders at different price levels based on their view of the market, creating a traditional order book. The market maker makes a profit by buying at lower prices and selling at higher prices. While this approach offers greater control over trade execution, it can suffer from low liquidity, higher spreads between the bid-ask price, and susceptibility to price manipulation. In AMM-based market making, a pool of liquidity is created on the AMM by passive Liquidity Providers (LPs) that act as a counter-party for all trades. The buy and sell prices are calculated based on a mathematical formula, and the AMM charges a small fee on all trades to incentivize the liquidity providers. AMM-based markets enable smaller traders to participate in the market, as they do not need to provide large amounts of capital to create an order book. AMM-based markets are more resistant to market manipulation since the price is determined by a mathematical formula. However, the liquidity providers are exposed to the risks of Impermanent Loss (IL) and smart contract security risks. The use of a fixed mathematical formula to determine the price of the contract can result in the price deviating from the spot price of the underlying asset, leading to arbitrage opportunities and other market inefficiencies\cite{daian2019flash}. PariFi is a decentralized perpetual protocol that aims to make trading on-chain derivatives more efficient for traders and liquidity providers. PariFi enables all user classes to gain exposure to an asset class without ever owning or interacting with the underlying asset. The assets can vary from cryptocurrencies, stocks, fiat, or commodities, as long as an on-chain price oracle for the asset is available. Traders can take long or short positions on these assets with leverage based on their risk profile. These trader positions are backed by depositing some collateral value, and the AMM provides them with leverage for a borrowing fee. The positions can be left open indefinitely without the need to roll them over\cite{he2023fundamentals}, as long as the collateral deposited is sufficient to cover the fees and any unrealized losses. Liquidity providers (LPs) can supply liquidity to the AMM to earn a share of the protocol revenue \cite{uniswap:1}. The majority share of the protocol revenue is distributed to the LPs for supporting the AMM, and a portion of this fee goes to the protocol treasury. A user can earn passive yield by providing liquidity to the AMM without having to deal with complex software and sophisticated tools for active market making. The protocol generates revenue from the open/close position fee, total borrowing fees, and liquidation fees. Apart from the fees, the protocol also generates revenue from trader losses; however, all trader profits are paid from the liquidity deposited in the protocol. PariFi aims to bring parity to traders and liquidity providers alike by balancing both sides of the trades—by maintaining low and attractive fees for traders while also ensuring that liquidity providers are fairly compensated for the risk of passive market making. We aim to achieve these goals by implementing novel features as listed below:

\begin{itemize}
    \item \textit{Liquidity Curve:} PariFi uses liquidity curves and on-chain oracles to create an adaptive pricing framework, which ensures market stability and predictability. It is a measure of market liquidity and is intended to have similar effects of the bid-ask price spread of an order-book based protocol. The framework uses parabolic curves to quote prices that account for the type of asset, market conditions and the total utilization of liquidity. This is described in more detail in Section \ref{sec:LiquidityCurve}.

\item \textit{Base and Dynamic Borrowing Fees:} The adaptive pricing framework introduces a system of base and dynamic borrowing fees, enabling the protocol to adjust fees based on prevailing market conditions. To dynamically modify these fees according to market conditions, the protocol employs a parabolic curve for the base borrowing fee and a sigmoid curve for the dynamic borrowing fee. This flexibility benefits both traders and liquidity providers by ensuring competitive fees for traders while fairly compensating liquidity providers for their contributions to the AMM. The different components of these fees and their formulas are explained in Section \ref{sec:BorrowingFees}, along with their mathematical models.

\item \textit{Improved Price Oracles:} The protocol implements a two-step process for order creation and settlement, coupled with multiple price oracles, to address challenges related to reliability, accuracy, and security. This approach mitigates risks such as front-running and price manipulation by updating the oracle price on-demand during settlement and allowing traders to set a maximum slippage. By utilizing multiple oracles and comparing their prices, the protocol ensures resilience against single-source data issues and selects the least favorable price to the trader within acceptable limits to further enhance fairness and security.

\end{itemize}

\textbf{Other features:}
\begin{itemize}
\item \textit{Auto-compounding LP vaults}
\item \textit{Meta Transactions (Gasless transactions)}
\item \textit{Limit Orders}
\item \textit{Risk Mitigation for markets}
\end{itemize}

\section{Liquidity Curve} \label{sec:LiquidityCurve}

PariFi employs an adaptive pricing framework that uses a parabolic liquidity curve to provide asset price quotes to traders. This U-shaped symmetric curve is utilized for quoting prices on both sides of the trade. Similarly, Curve Finance employs a hyperbolic curve for its constant product invariant to maintain liquidity balance.\cite{curve:1}

\vspace{0.25cm}The general function of a degree 2 parabola is

\begin{equation}
f(x) = ax^2 + bx + c
\end{equation}

with $a, b, c \in \mathbb{R}, a \neq 0.$

\vspace{0.5cm}
In our case, the liquidity curve is a function of utilization and is used to calculate the deviation of the quoted price from the actual price provided by the on-chain price oracle.

\subsection{Motivation}
In any order-book-based protocol, the bid-ask spread serves as the de facto measure of market liquidity. The bid-ask price spread is the difference between the buy price and the sell price of the asset quoted by the Market Maker (MM) in an order-book-based protocol/exchange. A trader accepts one of these prices based on whether they wish to buy or sell the asset. The spread increases as liquidity becomes more scarce or if the market conditions are not favorable to the market makers\cite{he2023fundamentals}. Market makers and professional traders who recognize imminent risk in the markets may also widen the difference between the best bid and the best ask they are willing to offer at a given moment.

In the case of an AMM-based perpetual protocol like PariFi, it is crucial to factor in the available liquidity, risk profile of the asset, and current market conditions when quoting prices. If the protocol fails to capture these factors in quoting prices, it would result in direct losses to the protocol and LPs. Hence, it is imperative to consider these points when quoting prices to traders. We achieve these goals by using the liquidity curve, which calculates the deviation or virtual spread from the oracle price based on the current Utilization (u) of the pool. We also use the constants Deviation Coefficient ($k_{\delta}$) and a Deviation Constant ($c_d$), which adjust the curve based on different parameters.

\subsection{The Curve} \label{ss:liquiditycurve}

The liquidity curve used by PariFi is a function of Utilization and Deviation and has the formula:

\begin{equation}
\delta = k_{\delta} \cdot u^2 + c_d
\end{equation}

where
\textit{
    \begin{itemize}
    \item $\delta$ = Deviation percentage
    \item $u$ = Current Utilization percentage
    \item $k_{\delta}$ = Deviation coefficient
    \item $c_d$ = Deviation constant
    \end{itemize}
    }

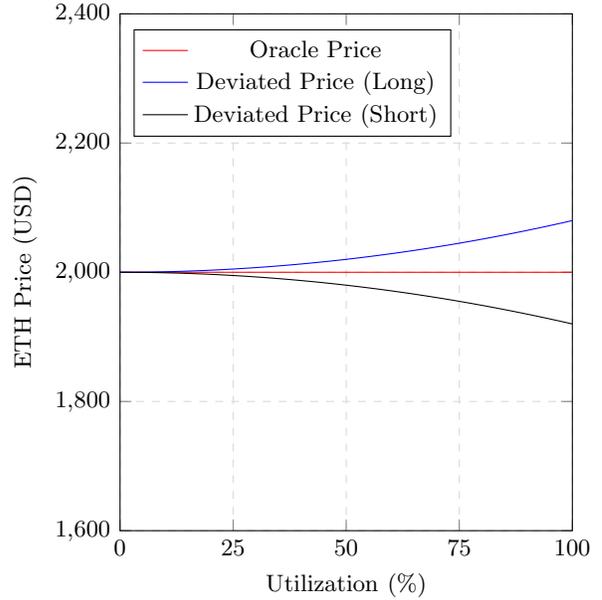
\begin{figure}[htbp]
\begin{tikzpicture}
\begin{axis}[
    title style={align=center},
    xlabel={\small Utilization (\%)},
    ylabel={\small ETH Price (USD)},
    tick label style={font=\small},
    xmin=0, 
    xmax=100, 
    ymin=1600, ymax=2400,
    xtick distance=25,
    legend pos=north west,
    grid=major,
    grid style={dashed, gray!30},
    width=0.46\textwidth, 
    height=0.37\textheight, 
    cycle list name=color list,
]

\addplot table [
    col sep=comma,
    x=Utilization,
    y=OraclePrice,
    ] {constant.csv};
\addlegendentry{\small Oracle Price}

\addplot table [
    col sep=comma,
    x=Utilization,
    y=DeviatedPriceLong,
    ] {constant.csv};
\addlegendentry{\small Deviated Price (Long)}

\addplot table [
    col sep=comma,
    x=Utilization,
    y=DeviatedPriceShort,
    ] {constant.csv};
\addlegendentry{\small Deviated Price (Short)}

\end{axis}
\end{tikzpicture}
\caption{\centering Utilization vs Price Deviation with constant price of ETH = 2000 USD and Deviation Coefficient $k_{\delta} = 0.0004$}
\label{fig:utilization_vs_deviation_constant_price}
\end{figure}

The formula models the relationship between liquidity and asset prices and is utilized to quote prices for both sides of the trade, ensuring pricing stability and predictability for traders. Figure \ref{fig:utilization_vs_deviation_constant_price} plots a graph of Utilization vs Deviated ETH Price when the price of ETH is constant at 2000 USD and the value of Deviation Coefficient ($k_{\delta}$) is $0.0004$. As utilization $u$ of the market increases, the deviated price quoted to the trader gradually moves away from the actual price received from the price oracle. The curve is used to control the steepness of the deviation. The impact of utilization on the deviation $\delta$ is adjusted using the deviation coefficient ($k_{\delta}$), as illustrated in Figure \ref{fig:utilization_vs_deviation_coefficient}.

\begin{figure}[htbp]
\centering
\begin{tikzpicture}
 \begin{axis}[
     title style={align=center},
     xlabel={\small Utilization (\%)},
     ylabel={\small ETH Price (USD)},
     tick label style={font=\small},
     xmin=0, 
     xmax=100, 
     ymin=2000, ymax=4000,
     ytick={2000,2500,3000,3500,4000},
     xtick distance=25,
     legend pos=north west,
     grid=major,
     grid style={dashed, gray!30},
     width=0.46\textwidth, 
     height=0.37\textheight, 
     cycle list name=color list,
]

\addplot table [
    col sep=comma,
    x=Utilization,
    y=OraclePrice,
    ] {table.csv};
\addlegendentry{\small Oracle Price}

\addplot table [
    col sep=comma,
    x=Utilization,
    y=DeviatedPriceLong,
    ] {table.csv};
\addlegendentry{\small Deviated Price (Long)}

\addplot table [
    col sep=comma,
   x=Utilization,
    y=DeviatedPriceShort,
    ] {table.csv};
\addlegendentry{\small Deviated Price (Short)}

\end{axis}
\end{tikzpicture}
\caption{\centering Utilization vs Price Deviation with historical ETH prices and Deviation Coefficient $k_{\delta} = 0.0004$}

\label{fig:eth_price_vs_utilization}
\end{figure}
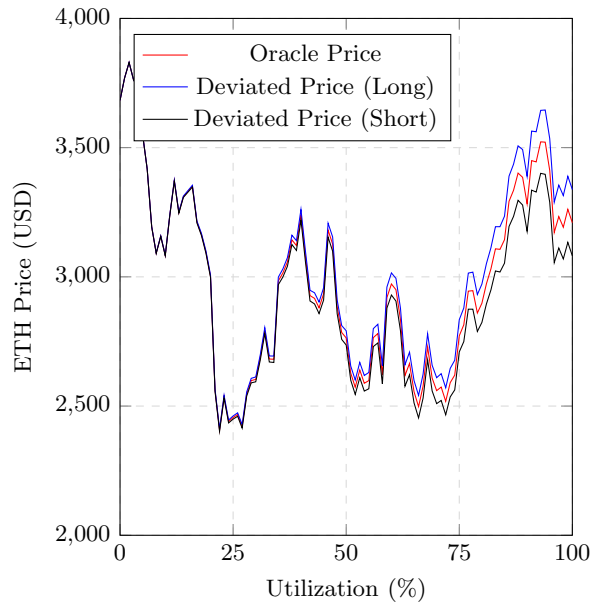

\subsection{Utilization ($u$)}

Utilization of the pool is a measure of how much liquidity is available to traders to open positions. When more liquidity is available, traders are encouraged to open positions by quoting prices that are very close to the actual market price. This is similar to the very low/negligible spread in the case of an order-book exchange. As more positions are opened by traders, liquidity is locked for these positions, reducing the total available liquidity and increasing the utilization. As utilization increases, it should become more expensive to borrowing funds for traders, and less favorable prices are quoted for the trades. This is similar to a high bid-ask spread in the case of an order-book exchange.

\subsection{Deviation Coefficient ($k_{\delta}$)}

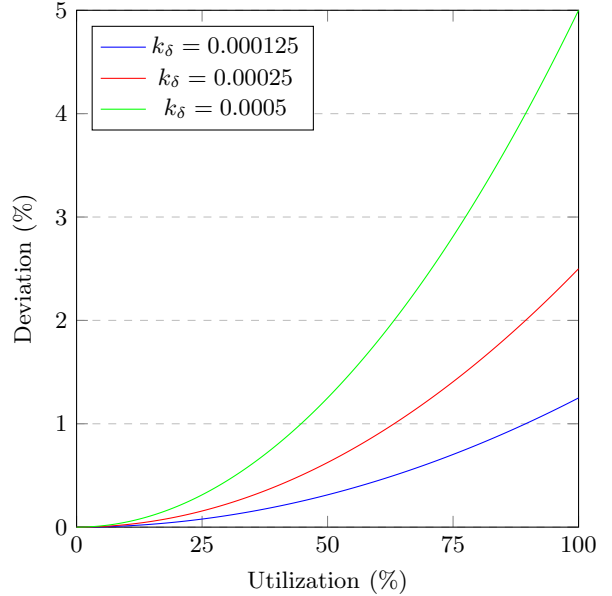
\begin{figure}[htbp]
\begin{tikzpicture}
\begin{axis}[
    width=0.5\textwidth, 
    height=0.37\textheight, 
    xlabel={\small Utilization (\%)},
    ylabel={\small Deviation (\%)},
    tick label style={font=\small},
    font=\small,
    xmin=0, xmax=100,
    ymin=0, ymax=5,
    xtick={0,25,50,75,100},
    ytick={0,1,2,3,4,5},
    legend pos=north west,
    ymajorgrids=true,
    grid style=dashed,
]

\addplot[
    color=blue,
    smooth,
    ] table [x=Utilization, y=Deviation_dc_0.000125, col sep=comma] {devi.csv};
\addlegendentry{\small $k_{\delta}=0.000125$}

\addplot[
    color=red,
    smooth,
    ] table [x=Utilization, y=Deviation_dc_0.00025, col sep=comma] {devi.csv};
\addlegendentry{\small $k_{\delta}=0.00025$}

\addplot[
    color=green,
    smooth,
    ] table [x=Utilization, y=Deviation_dc_0.0005, col sep=comma] {devi.csv};
\addlegendentry{\small $k_{\delta}=0.0005$}
\end{axis}
\end{tikzpicture}

\caption{\centering Impact of Utilization on Deviation for different values of Deviation Coefficient $k_{\delta}$ }
\label{fig:utilization_vs_deviation_coefficient}
\end{figure}

The Deviation Coefficient ($k_{\delta}$) is a constant set by the protocol based on the risk profile of the asset and market conditions. As the value of $k_{\delta}$ increases, the curve becomes steeper, and deviation increases more rapidly with an increase in utilization. Assuming that the market conditions are the same, a low value for this constant means that the asset is more stable, while a high value means that the asset is more volatile. Figure \ref{fig:utilization_vs_deviation_coefficient} is a graph plotted with Utilization($u$) vs Deviation($\delta$) for different values of the deviation coefficient ($k_{\delta}$). As we can see, the deviation increases more rapidly with utilization for higher values of the coefficient and vice versa.

\subsection{Deviation Constant ($c_d$)}
The Deviation Constant ($c_d$) is used to set the minimum deviation for the curve for a specific asset. Initially, the value of this constant is set to zero, but can be fine-tuned based on the risk profile of the asset and market conditions.

\section{Base and Dynamic Borrowing Fee} \label{sec:BorrowingFees}

In the case of an AMM-based perpetual protocol, when a trader opens a leveraged position, they are essentially borrowing money from the liquidity providers (LPs) of the AMM. The LPs expect a return on the capital that is lent to traders. The cost of borrowing capital for a leveraged trade should vary with the current market conditions and available liquidity. The AMM should be fairly compensated for the risks associated with lending liquidity, as all trader profits are a direct loss to the AMM. Therefore, the borrowing fees should reflect this added risk. The borrowing fees are expected to be higher than the lending rate of low-risk over-collateralized web3 lending markets like Aave\footnote{https://aave.com/} and Compound\footnote{https://compound.finance/} as these markets are not exposed to this risk.

The protocol has been designed to ensure that the fees reflect the points mentioned above, providing fairness for both traders and liquidity providers. The total borrowing fees consist of the Base Borrowing Fees ($F_b$) and the Dynamic Borrowing Fees ($F_d$). The total borrowing fee for a position is equal to:

\begin{equation}
 TBF =  F_b + F_d 
\end{equation}

where

\textit{
    \begin{itemize}
        \item $TBF$ = Total borrowing fee
        \item $F_b$ = Base borrowing fee
        \item $F_d$ = Dynamic borrowing fee
    \end{itemize}
    }

The Base Borrowing fee is charged for all trades, while the Dynamic borrowing fee is charged only to one side, depending on the Market Skew of the open positions. These components are explained in detail below.

\subsection{Base Borrowing Fee}

The Base Borrowing Fee ($F_b$) is the fee charged to both sides of the trade for borrowing liquidity from the AMM. The fee is different for each asset and is calculated using a Parabolic Curve similar to the liquidity curve explained in Section \ref{sec:LiquidityCurve}. The base borrowing fee depends on factors such as the risk profile of the asset, current market volatility, and liquidity utilization.
The Base Borrowing Fee ($F_b$) is calculated using the formula

\begin{equation}
F_b = k_b \cdot u^2 + c_b
\end{equation}

where

\textit{
    \begin{itemize}
    \item $F_b$ = Base borrowing fee
    \item $k_b$ = Base coefficient
    \item $u$ = Utilization percentage
    \item $c_b$ = Base constant
    \end{itemize}
}

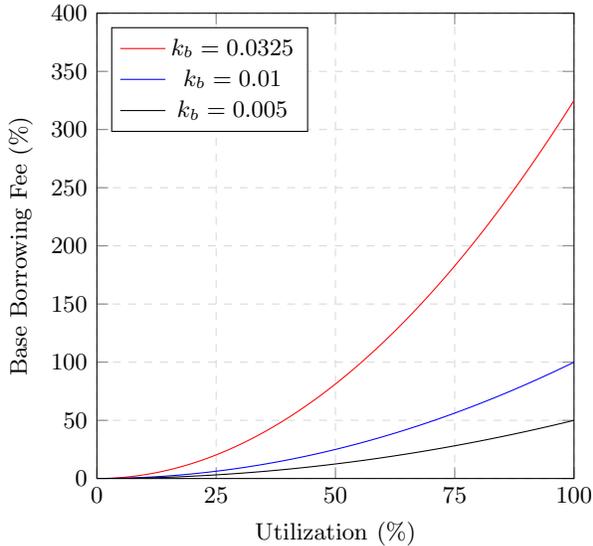
\begin{figure}[htbp] 
\centering
\begin{tikzpicture}
\begin{axis}[
    xlabel={\small Utilization (\%)},
    ylabel={\small Base Borrowing Fee (\%)},
    tick label style={font=\small},
    xmin=0, xmax=100,
    ymin=0, ymax=400,
    xtick={0,25,50,75,100},
    ytick={0,50,100,150,200,250,300,350,400},
    legend pos=north west,
    grid=major,
    grid style={dashed, gray!30},
    width=0.48\textwidth, 
    height=0.34\textheight, 
    cycle list name=color list,
]

\addplot table [
    col sep=comma,
    x=Utilization,
    y={Base Fee (b = 0.0325)},
    ] {basefee.csv};
\addlegendentry{\small $k_b=0.0325$}

\addplot table [
    col sep=comma,
    x=Utilization,
    y={Base Fee (b = 0.01)},
    ] {basefee.csv};
\addlegendentry{\small $k_b=0.01$}

\addplot table [
    col sep=comma,
    x=Utilization,
    y={Base Fee (b = 0.005)},
    ] {basefee.csv};
\addlegendentry{\small $k_b=0.005$}

\end{axis}
\end{tikzpicture}
\caption{\centering Utilization vs Base Borrowing Fees for different values of Base Coefficient ($k_b$)}
\label{fig:base_borrow_fees_vs_utilization}
\end{figure}

The Base Coefficient ($k_b$) is used to control the steepness of the Borrowing Fee curve. A lower value of the base coefficient results in a small increase in the base borrowing fee with utilization, while higher values of the coefficient result in rapid increases in the fee as utilization increases. This relationship is represented in Figure \ref{fig:base_borrow_fees_vs_utilization}, where a line graph of Utilization vs Base Fee is plotted for different values of the Base Coefficient. As we can see, when $k_b=0.005$, the borrowing fee curve increases slowly and gradually as utilization increases. However, when $k_b=0.0325$, the borrowing fee increases rapidly as utilization increases.

It is essential to choose the base coefficient for an asset wisely, depending on the stability and risks associated with the market-making of the asset.

The Base constant ($c_b$) is initially set to 0 and can be modified later by the protocol based on market activity and governance. When the value of this constant is 0, the base borrowing fee starts from zero. It can be used to set a minimum amount of borrowing fees so that the curve is shifted up by this value.

\subsection{Dynamic Borrowing Fee} \label{ss:dynamic_borrowing_fee}

In an order-book-based perpetual protocol, a Funding Rate is used to keep the mark price close to the index price and to balance the markets. Funding payments are paid by traders that move the mark price away from the market price and earned by traders that move the price close to the market price\cite{he2023fundamentals}. However, in the case of an AMM-based solution like PariFi, as prices are quoted by the price oracles using the liquidity curve, we do not have a funding rate. However, to balance the market skew of active positions and to deleverage risks to the protocols, PariFi introduces a Dynamic Borrowing Fee.

The Dynamic Borrowing fee is used to balance the skew in markets and is charged to only one side depending on the imbalance in Open Interest (OI) of long and short positions. This fee is paid by the side which has the higher OI to the AMM, as the AMM is the Market Maker here on behalf of the LPs that take on the risk by providing liquidity. This helps keep the overall fees low, as all interest earned goes to the LP, resulting in a higher yield for liquidity provided. This also benefits the traders as they can take advantage of lower borrowing fees.

The Dynamic Borrowing Fee is a Sigmoid function and is calculated using the formula:

\begin{equation}
F_d = \cfrac{M\cdot(1-\cfrac{1}{e^{k\sigma}})}{(1+\cfrac{1}{e^{k\sigma}})}
\end{equation}

\vspace{0.5cm}

where

\textit{
\begin{itemize}
    \item $F_d$ = Dynamic Borrowing fee
    \item $M$ = Maximum dynamic borrowing fee
    \item $\sigma$ = Market Skew
    \item $k$ = Constant used to determine the steepness of the sigmoid function
\end{itemize}
}

\vspace{0.25cm}
The same formula can be represented in a simplified form as 

\begin{equation}
F_d = \frac{M\cdot (1 - e^{-k\sigma})}{(1 + e^{-k\sigma})}
\end{equation}

\vspace{0.5cm}
The Dynamic Borrowing Fee ($F_d$) depends on the market skew $\sigma$, a constant $k$ that controls the steepness of the curve, and the maximum value of the dynamic borrowing fee $M$.

\vspace{0.5cm}

The market skew ($\sigma$) is calculated using the open interest of long and short positions as follows:

\vspace{1.5cm}

\begin{equation}
\sigma = \frac{|L - S|}{P}
\end{equation}

where

\textit{
\begin{itemize}
    \item $\sigma$ = Market skew
    \item $L$ = Open interest of long positions
    \item $S$ = Open interest of short positions
    \item $P$ = Total pool value
\end{itemize}
}

\begin{figure}
    \centering

\begin{tikzpicture}
\begin{axis}[
    width=0.46\textwidth, 
    height=0.37\textheight, 
    xlabel={\small Market Skew (\%)},
    ylabel={\small Dynamic Borrowing Fee (\%)},
    font=\small, 
    xmin=0, xmax=100,
    ymin=0, ymax=500,
    xtick={0,25,50,75,100},
    ytick={0,100,200,300,400,500},
    legend pos=north west,
    ymajorgrids=true,
    grid style=dashed,
]

\addplot[
    color=blue,
    mark=line,
    ] table [x=Skew_Percent, y=DynamicBorrowFee_0.0125, col sep=comma] {dynamics.csv};
\addlegendentry{\small $k=0.0125$}

\addplot[
    color=red,
    mark=line,
    ] table [x=Skew_Percent, y=DynamicBorrowFee_0.0225, col sep=comma] {dynamics.csv};
\addlegendentry{\small $k=0.0225$}

\addplot[
    color=green,
    mark=line,
    ] table [x=Skew_Percent, y=DynamicBorrowFee_0.0325, col sep=comma] {dynamics.csv};
\addlegendentry{\small $k=0.0325$}

\end{axis}
\end{tikzpicture}
    \caption{\centering Market Skew ($\sigma$) vs Dynamic borrowing fees ($F_d$) for different values of $k$ }
    \label{fig:skew_dynamic_fees_k}
\end{figure}
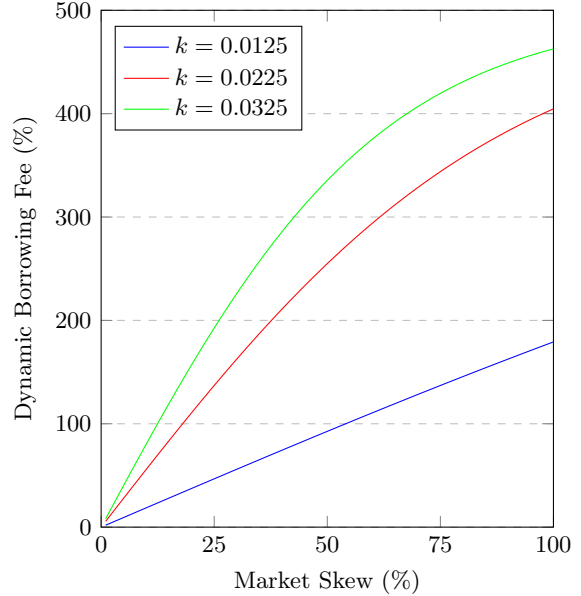

Figure \ref{fig:skew_dynamic_fees_k} plots a line graph of Market skew ($\sigma$) vs Dynamic Borrowing Fee ($F_b$) for different values of the steepness constant $k$. For lower values of $k$, the dynamic borrowing fees increase almost linearly with the market skew. However, as the value of $k$ increases, the dynamic borrowing fees increase rapidly even for a small increase in the market skew. Stable assets should be configured to have low values of $k$, while very risky and volatile assets should have a higher value of $k$.

\section{Improved Price Oracles}

Oracles are a bridge between the real world and blockchain networks. Price oracle smart contracts are a type of external data source that provides up-to-date information on the price of assets such as cryptocurrencies, stocks, commodities, or fiat currencies for use in decentralized applications. Many protocols use price oracles for executing smart contracts that involve asset trading or price-dependent decisions. Using price oracles can present several challenges that protocols must address to ensure the reliability, accuracy, and security of the data provided. Other risks, such as front-running (though external to the price oracles themselves), also pose a threat to the protocols. Front-running refers to the practice of taking advantage of advanced knowledge of real-world data or pending transactions to gain an unfair advantage in a trade. A lot of research has been put into transaction reordering and front-running of Price Oracles in decentralized markets \cite{daian2019flash}. Due to the nature of how blockchain smart contracts work in an adverse environment, the trader always can take advantage of information asymmetry and exploit the price difference for an unfair advantage \cite{chainlink:1} \cite{daian2019flash}.

PariFi uses a combination of techniques to address these challenges. The protocol uses a two-step process for Order creation and Order settlement. The execution price is the oracle price at the instant the transaction is settled. Traders can set a maximum slippage between their submitted price and the execution price. If the price deviates more than the slippage from the instance the order is submitted until the time it is settled, the transaction is reverted. During order settlement, the protocol requests an oracle price update on-demand \cite{pyth:1} from a network of price aggregators, so the execution price is always the latest price available for consumption. There are additional checks on the timestamp these feeds were last updated, and any outdated price is rejected.

PariFi uses multiple price oracles to settle the order. This ensures that the protocol is not prone to price manipulation or incorrect feed at a single source of data. During order settlement, the price from both data sources is compared, and if the deviation is above the configured threshold limit, then the transaction reverts. In case the deviation between both the oracle prices is more than the minimum acceptable limit but less than the threshold limit, then the protocol uses the price which is least favorable to the trader and more favorable to the protocol. 

\section{Other Features}

\subsection{Auto-compounding LP vaults}

Auto-compounding LP Vaults play an essential role in enhancing the perpetual protocol's efficiency and user experience. Built on the EIP-4626 standard, these vaults automatically compound the fees earned by the Automated Market Maker (AMM) over time, providing liquidity providers with a passive income strategy that eliminates manual claims for rewards or fees.

The PariFi protocol offers two types of vault assets: stable and volatile assets. This distinction allows liquidity providers to select their desired risk profile when engaging with the protocol. Profits and fees generated from trading activities are directly credited to the vault, increasing the value of vault tokens over time. It is crucial to note that the protocol's losses, such as trader profits, are also paid from this vault, ensuring a balanced ecosystem that aligns the interests of traders and liquidity providers.

By implementing auto-compounding LP Vaults, PariFi aims to simplify the process for liquidity providers while maximizing their potential returns in a user-friendly and efficient manner.

\subsection{Meta Transactions (Gasless transactions)}

Meta Transactions, or gasless transactions, represent a key feature that enhances the protocol's functionality and user experience. The approach allows users to pay for gas fees using ERC20 collateral tokens, eliminating the need for multiple token approvals and streamlining the transaction process.

A decentralized network of Relayers\cite{gelato:1} submits transactions on behalf of users, ensuring a seamless and efficient user experience. The protocol relies on cryptographic signatures to verify user wallets and employs secure implementations of the finalized EIP standards EIP-712 and EIP-2771 to maintain the highest levels of security and trust.

Users are also able to set a deadline, after which a transaction cannot be executed, adding an extra layer of control and flexibility to the trading process.

By integrating meta-transactions into the protocol, PariFi aims to offer a more accessible and user-friendly trading environment that caters to the evolving needs of both traders and liquidity providers in the decentralized finance ecosystem.

\subsection{Limit Orders}

The integration of Limit Orders, including Stop Loss, Take Profit, and Create Order functionalities, significantly enhances the versatility and adaptability of the PariFi perpetual protocol. These features empower traders to better manage their risk and optimize their trading strategies in a dynamic market environment.

Stop Loss orders enable traders to set a predetermined price level at which their position will be automatically closed, minimizing potential losses in case the market moves against their prediction. Conversely, Take Profit orders allow traders to lock in profits by closing their position once the market reaches a specific price level in their favour.

Create Order functionality offers traders the ability to place orders at specific price points, enabling more strategic and precise execution of trades. This feature is particularly useful in volatile markets, where rapid price fluctuations can present significant opportunities for profit.

\subsection{Risk Mitigation for Markets}

PariFi has configurable limits per market that allows the protocol’s governance to set customized risk parameters for every asset, to ensure a secure and efficient trading environment that caters to the needs of both traders and liquidity providers.

The protocol can adjust risk parameters such as maximum open interest, leverage, and exposure for each asset. This implementation effectively mitigates potential risks and upholds a balanced trading environment, safeguarding the interests of all market participants. Asset volatility risks are also controlled using the fields Base Coefficient $k_b$ and Base Constant $c_b$ of the Base Borrowing Fee and the steepness constant $k$ of the Dynamic Borrowing Fee, explained in Section \ref{sec:BorrowingFees}. Liquidity risks can be controlled by configuring the Deviation coefficient $k_{\delta}$ and Deviation constant $c_d$ of liquidity curve, which is explained in Section \ref{sec:LiquidityCurve}. 

Traders benefit from the ability to adjust their strategies according to predefined risk tolerances, while liquidity providers gain from controlled exposure to specific assets. This approach ensures precise and accessible risk management, fostering a sustainable and stable market experience for all users.

\section*{Disclaimer}
This paper is for general information purposes only. It does not
constitute investment advice or a recommendation or solicitation to
buy or sell any investment and should not be used in the evaluation
of the merits of making any investment decision. It should not be
relied upon for accounting, legal or tax advice or investment recommendations. This paper reflects current opinions of the authors
and is not made on behalf of ETHA Labs, PariFi, or their
affiliates and does not necessarily reflect the opinions of ETHA Labs, PariFi, their affiliates or individuals associated with them.
The opinions reflected herein are subject to change without being
updated.

\section*{License}

This work is licensed under the Creative Commons Attribution-NonCommercial-NoDerivatives 4.0 International License. To view a copy of this license, visit http://creativecommons.org/licenses/by-nc-nd/4.0/ or send a letter to Creative Commons, PO Box 1866, Mountain View, CA 94042, USA.

\vspace{0.25cm}
\centering
\ccbyncnd

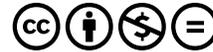
\end{document}